\documentclass[aps,prd,superscriptaddress,eqsecnum,amsfonts,showpacs,epsfig]{revtex4}
\usepackage{graphicx}

\newcommand{\be}{\begin{equation}}
\newcommand{\ee}{\end{equation}}
\newcommand{\bea}{\begin{eqnarray}}
\newcommand{\eea}{\end{eqnarray}}

\newcommand{\nn}{\nonumber}
\newcommand{\ep}{i\epsilon}
\newcommand{\om}{\omega}

\begin{document}

\title{Towards the finite quantum field theory}
\author{V. \v{S}auli:       }
\affiliation{Department of Theoretical Physics, NPI Rez near Prague, Czech Academy of Sciences}

\begin{abstract}
  In this study, we propose a novel regularization/renormalization scheme that utilizes an auxiliary Feynman parameterization. This approach is employed to align a specified loop diagram with a designated unit of the form $1=\lambda/\lambda$. 
Within   the proposed regularization technique, we formulate the standard renormalization scheme
and demonstrate  conditions under which it yields symmetry preserving results.
It is demonstrated that its minimal form yields renormalized diagrams that are equivalent to those of the dimensional renormalization scheme, with the exception of their counterterms. 
Furthermore,  a novel procedure for taking the soft limit $\lambda\rightarrow 0$, where a properly defined order of computational actions provides the field theory completely finite, is presented.The qualitative and quantitative distinctions between this approach and the standard scheme are highlighted.
Both schemes are elucidated  in  the scalar model in 3+1D for pedagogical reasons. Subsequently,  the  proposed schemes are applied to the  Standard Model at one loop level,  e.g. we calculate photon and gluon polarizations.    QCD effective charge is calculated in the finite QCD, exhibiting a clear evidence that the finite QCD is not ruled out by experiment.  In the final section, we offer a concise discussion on the softening of anomalies and the treatment of overlapping divergences, accompanied by illustrative examples.

\end{abstract}
\maketitle

%%%%%%%%%%%%%%%%%%%%%%%%%%%%%%%%%%%%%%%%%%%%%%%%%%%%%%%%%%%%%%%%%%%%%%%%%%%%%%%%%%%%%

\section{Introduction}

 Renormalization is the procedure that should be performed to obtain a meaningful comparison between quantum field theory predictions and experiments. The associated recipe  contains the renormalized Green's functions and unphysical counterterm parts at the end. When applied to processes describing experiments, the former are used as a 
 building blocks for matrix transitions, while the latter contain  divergences, that  more closely resemble  the way they have been calculated, rather then subject of renormalization.  Irrespective of the value of couplings in game, the counterterms become an expansion in powers of infinite regulators , t'Hooft $\epsilon_d^{-1}$ in Dimensional regularization or Pauli-Villars cutof $\Lambda$ to name a few.  
A notable concordance exists between theoretical predictions and experimental observations concerning the muon magnetic moment,on the muon magnetic moment \cite{mu21,mu22,mu23}, and the precise electroweak and QCD precise fits \cite{Z-shape1,Z-shape2,Z-shape3,Z-shape4} .
The interrelation of both \cite{both,both2}, recently based on multi-loop Feynman diagrams, represents the test of the Standard Model within the renormalization program as its seemingly unavoidable part.

In the aftermath of publication of the paper \cite{THVE1972},  it was determined that the most economical approach to address the renormalization problem is to utilize the method of dimensional renormalization/regularization (DIMR).The old-fashioned approaches \cite{VHP, BPHZ1,BPHZ2,BPHZ3,BPHZ4} were found  to be cumbersome when dealing with overlapping divergences in multiloop calculations.
 However, the leading role of the Pauli-Villars regularization method in the development of the DIMR scheme was indispensable in the case of chiral anomalies \cite{ABJ1}, \cite{ABJ2}. Despite the demonstrated efficacy of the Standard Model in being algebraically renormalizable, i.e., without the use of any regularization, the DR scheme remains the most effective approach to renormalize quantum corrections.

 The mass hierarchy problem was attributed to the theoretical introduction of supersymmetry, which was itself motivated by a naively quadratically divergent diagram. Following the experimental confirmation of the relatively light Higgs boson \cite{higgs1},\cite{higgs2} and practical exclusion of supersymmetry at currently accessible energies, 
 this is still the self-energy in the Higgs boson propagator, which occasionally gives rise to a suspicious friction between bare and renormalized parameters of the Standard Model.  Consequently, the corresponding problems of naturalness may be ill-defined in perturbation theory and may not be problematic at all, as asserted in Ref. \cite{SR2019}.   To this point, the DIMR has initiated the treatment of the quadratic divergence as if it were logarithmic. Both divergent diagarams are then proportional to single inverse power of $\epsilon_d$ parameter.
  In the present study, we make an advance in the state of the field by introducing a well-defined regularization technique. This technique effectively renders both logarithmic and quadratic divergences into finite forms. However, it should be noted that this approach involves a trade-off, as it employs a more intricate calculation technique compared to the DIMR, necessitating an additional Feynman parameterization.

 The study of hadrons at low energies is certainly beyond the framework of perturbation 
theory.   However, the formalism of the Dyson-Schwinger equations  \cite{DSE0,DSE1,DSE2,DSE3,DSE4} is a special example
 where the perturbative technique can be combined with nonperturbative framework  in practice \cite{DSE4,ja2020,MESA2020,ja2022a,PAFRSA2023,ja2022b,ACS2023,ja2023,HPW2023}.  
 In fact,  non-perturbative phenomena such as gluon and  quark mass generation are governed by quantum loops of strongly coupled QCD. 
Only after careful removal of ultraviolet (UV) singularities from the equations, the  meaningful predictions for hadron properties can be obtained. 
 The presented regularization technique in combination with integral representations keeps  non-perturbative calculations free from problematic momentum integrations  without destroying the predictive power of the theory.  The development of  the  Finite Regularization Scheme presented in this paper, was highly motivated by these studies and  may be used for this purpose in the near future.

  In this paper we offer the regularization$\&$renormalization scheme that 
 preserves all required symmetries in at least one loop level. Unlikely to t'Hooft dimensional regularization/renormalization scheme
 gauge invariance is not automatically ensured for naively regularized amplitudes.
  Nevertheless, it is not difficult to formulate conditions and computational rules that ensure gauge invariance of the presented renormalization scheme. 

  In the next section we introduce the technical details and apply them to an instructive pedagogical model in section 3.
  Further application is done in Sections 4 and 5, where we present the calculation of the photon  (gluon) polarization function in  SM (QCD).
 We present the Standard Renormalization Scheme and discuss the question of whether Finite Regularization/Renormalization Scheme is appropriate in each individual case. 
  Conceptual features and differences to conventional schemes are shown for the ABJ anomaly in Sec. 6. 
We also show how to disentangle with overlapping divergences and derive the result for the two-loop sunset diagram in detail in Sec. 7.
 We will  work in 3+1D Minkowski space, or equally in the 4D Euclidean space and 
 do not  continue to non-integer dimensions. This naturally avoids the problems with intrinsically integer 
 dimensional object like Levicivita pseudotensor $\epsilon_{\alpha\beta\gamma\delta}$ or $\gamma_5$ matrix. 

 \section{Definitions and main theorem}

The Feynman parameterization serves as a primary step, with most established RSs, including the t'Hooft-Veltman DR scheme and the algebraic BPHZ scheme, relying heavily on its implementation.   A critical aspect of the Feynman parameterization that has been overlooked is its ability to make the naked theory finite by strategically changing the order of actions.  This is achieved by a two-step procedure. The first step involves the standard practice of making all momentum space integrals finite prior to integration. This is achieved by transferring all divergences into the Feynman parameter integral in advance. This preliminary step allows the subsequent definition of a meaningful renormalization scheme: The Standard Renormalization Scheme (SRS)

 In order to achieve the aforementioned Finite Regularization/Renormalization Scheme (FRRS), the second step must be followed. This step is predicated on the specific selection of a regulator that vanishes.  Subsequently, the divergences, which are incorporated into the Feynman parameter integral, are canceled out.
 This introduces an element of arbitrariness, which is tolerable provided it results in a finite bare Lagrangean. 
  
To elucidate this methodology, we proceed to illustrate its application in the context of one-loop diagrams. 
Utilizing the Feynman parameterization, it is possible to transform any tensor integral into a scalar integral of the following form:
\bea \label{tch1}
i\int \frac{d^4 k}{(2\pi)^4}\frac{k^{\mu_1}...k^{\mu_j}}{[k^2+k\cdot p-M^2+\ep]^{\alpha}}\, 
=\prod_{m=0}^{j-1}\frac{1}{(j-m-\alpha)}\frac{\partial}{\partial p_{\mu_m}} 
\int \frac{d^4 k}{(2\pi)^4}\left(\frac{-\lambda_F^2}{-\lambda_F^2}\right)^f\frac{i}{[k^2+k\cdot p-M^2+\ep]^{\alpha-j}}\, \, \, ,
\eea
where we have inserted the unit with arbitrary  dummy parameter $\lambda_F$ for which we choose the dimension of  $mass$ for convenience.
In what follows, the  Feynman parametric integral  in the form
\be \label{fejn}
 \frac{1}{a^{\gamma_1}}\frac{1}{b^{\gamma_2}}=\frac{\Gamma(\gamma_1+\gamma_2)} {\Gamma(\gamma_1)\Gamma(\gamma_2)} 
 \lim_ {\epsilon_z\rightarrow 0}\int_{\epsilon_z}^1 d z \frac{z^{\gamma_1-1}(1-z)^{\gamma_2-1}}{[a z+b (1-z)]^{\gamma_1+\gamma_2}} \, ,
\ee
is applied once again, such that the  Eq. (\ref{tch1}) is equivalently written like 
\be
\prod_{m=0}^{j-1}\frac{1}{(j-m-\alpha)}\frac{\partial}{\partial p_{\mu_m}} \lim_ {\epsilon_z\rightarrow 0}\int_{\epsilon_z}^1 d z  \int \frac{d^4 k}{(2\pi)^4}
\frac{i \beta(f,\alpha-1) z^{(f-1)}(1-z)^{(\alpha-j-1)}(-\lambda_F^2)^f }{[(k^2+k\cdot p-M^2+\ep)z-\lambda_F^2(1-z)]^{\alpha-j+f}}\, \, \, ,
\ee
where  beta functions $\beta(\gamma_1,\gamma_2)$ is shorthand notation for the three gamma functions prefactor
 that appears in front of the limit at rhs. of Eq. (\ref{fejn}).  The choice of  $f$ is made such that $\alpha-j+f>2$ ensure  the momentum integral  convergent. A practical choice   $\alpha-j+f=3$
can be made, since integer values are always appreciated in multiloop calculations.

Performing the momentum integration gives
 \be \label{tch3}
\prod_{m=0}^{j-1}\frac{1}{(j-m-\alpha)}
\frac{\partial}{\partial p_{\mu_m}}
\lim_{\epsilon_z\rightarrow 0}
\frac{\Gamma(a-j+f-2)}{(4\pi)^2\Gamma(f)\Gamma(a-j)}
\int_{\epsilon_z}^1 d z  \frac{(-1)^{(f+1)}\lambda_F^{(2f)} (1-z)^{(\alpha-j-1)}}
{z^{(\alpha-j+1)}[p^2-M^2-\lambda_F^2\frac{(1-z)}{z}+\ep]^{\alpha-j+f-2}}\, \, ,
\ee  
where all gamma functions are finite. All potential UV divergences, if present, have been transported into the single Feynman integral.

The proposed RS has been developed for the Standard Model (and its extensions) quantized in renormalizable gauges. In this case, the highest degree of divergence that occurs for individual Feynman diagrams is two. In the expression (\ref{tch3}), the  real parameter $\lambda_F$ can take any  value. 
 To ensure the Feynman parameterization valid, the regulator $\epsilon_z$ must  be set to zero at the end of the calculation.
  In the event that a finite value is maintained for the parameter $\lambda_F$, 
  then those terms involving negative power of the regulator should be sent to the renormalization constant  of the quantized   theory.
   It will be demonstrated  that the  for the finite parameter  $\lambda_F$ one gets at least a single  pole of the 
 form $\epsilon_z^{-1}$ for quadratically divergent diagram, while the term   
$\ln\epsilon_z$ appears for  log divergent diagram.

 Let us consider the the following example of the scalar bubble integral:
\be \label{tchunko}
b(q^2,m^2)=i\int \frac{d^4 k}{(2\pi)^4}\frac{1}{[k^2-m^2+\ep][(k-q)^2-m^2+\ep]}\, .
\ee
can be calculated bu putting   $\alpha=2$ in the Eq. (\ref{fejn}) and hence it is enough to take $f=1$ for making the momentum integration finite.
Explicitly written:    
\bea
b(q^2,m^2)&=&i \int \frac{d^4 k}{(2\pi)^4}\int_0^1 d x \frac{-\lambda_F^2}{[-\lambda_F^2][k^2 x+(k-q)^2 (1-x)-m^2+\ep]^2}
\nn \\
&=&i  \int \frac{d^4 k}{(2\pi)^4}\int_0^1 d x L \int _{\epsilon _z}^{1} dz \frac{-2 z \lambda_F^2}{[(k^2 +q^2 x (1-x) -m^2) z -\lambda_F ^2(1-z)+\ep]^3} 
\nn \\
&=&\frac{L}{(4\pi)^2}\int_0^1 d x \int _{\epsilon _z}^{1} dz \frac{\lambda_F^2}{z [(q^2 x (1-x) -m^2 +\lambda_F^2 )z-\lambda^2_F+\ep]}
\nn \\
&=& \frac{L}{(4\pi)^2}\int_0^1 d x \left[\ln\frac{J-\ep}{\mu^2}
-\ln\frac{J-\lambda_F^2 +\lambda_F^2 \epsilon_z^{-1}}{\mu^2}\right] \,  ,
\label{dvojka}
\eea
where shorthand notation $J=-q^2 x (1-x)+m^2 $  is used, $\mu^2$ is an arbitrary renormalization scale and  where $\hat{L}$ stands  shortly for the limiting  symbol
\bea
 \hat{L} f=\lim_{\epsilon_z\rightarrow 0} f \, 
\eea
 where $f$ is a testing function.

 \subsection{Standard perturbative renormalization}
 
It is reasonable, although not unique, to posit that the structure above dictates which term is designed to be the renormalized one and which should belong
to the  counterterm $\delta$.  Of course, the standard renormalization scheme can be easily defined only  for the finite variable  $\lambda_F$ since 
then for vanishing $\epsilon_z$ one can neglect $J$ against the ''infinite'' term  $\lambda_F /\epsilon_z$ in the second log.   
In this limit one can  split the above result as the following  
\bea  \label{treti}
b(q^2;m^2)&=&b_R(q^2;m^2)+\delta_b
\nn \\
b_R(q^2;m^2)&=&\frac{1}{(4\pi)^2} \int_0^1 dx \ln\frac{-q^2 x (1-x)+m^2-\ep}{\mu^2}\, ,
\nn \\
\delta_b&=&\frac{\hat{L}}{(4\pi)^2}\ln\frac{\lambda_F^2}{ \epsilon_z\mu^2} \,.
\eea
Choosing  $\delta_b$ be a counterterm part provides 
the renormalized  part $b_R$    identical to  
the minimal subtraction  dimensional renormalization scheme (DIMR) .

 It is evident that the outcome of a given regularization scheme can be replicated in other regularization or renormalization schemes,
 provided that both schemes are applicable. 
The presented regularization scheme does not incorporate a momentum loop-dependent regulator, which is why a particularly straightforward relation with the DIMR is highly anticipated. Substantial changes come in the limit $\lambda_F\rightarrow 0$.

\subsection{($Z=1$) Finite renormalization/regularization scheme}

It is important to note that the case of $\lambda_F \simeq \epsilon_z$ as well as a complex valued  $\lambda_F$ is not excluded from the  set of permissible parameters.
 To that end, let us  consider the regularized equation (\ref{dvojka}) and perform the double limit  $\lambda_F,\epsilon_z \rightarrow 0$ ensuring the following rate is finite
\be
\lim_{\epsilon_z\rightarrow 0} \frac{-\lambda_F^2}{\epsilon_z}=\mu_F^2 \, .
\ee

Consequently, it is evident that the interchange of limits and integrations yields a finite yet ambiguous result:
\be  \label{diminish}
b(q^2;m^2)=\frac{1}{(4\pi)^2} \int_0^1 dx \ln\frac{-q^2 x (1-x)+m^2-\ep}{-q^2 x (1-x)+m^2+\mu_F^2}\, .
\ee
 
   Note that some care care should be taken  when performing this limit to avoid violation of Unitarity. For this purpose, the principal value of logarithmic function 
 was taken  in order to get rid of unwanted rise of the absorptive part at the unphysical threshold $2(m+\mu_F)$. Alternatively, this   can be  achieved formally by taking the  $\lambda$ complex and infinitesimaly small in order to cancel the Feynman $i\epsilon$.

In the following text we will use the shorthand notation
  \be  \label{FRS}
\mbox{Lg}(J;\mu_F)=\ln{[-J]}-Ln{[-J+\mu_F^2]}  \, 
\ee        
to preserve standard receipt for  analytical continuation to the timelike momentum.

For a spacelike momenta arguments $-J>0$ and for positive scale $\mu_F^2$ , the function \ref{FRS} turns to be a standard log
\be  
\mbox{Lg}(J;\mu_F)=\ln\frac{-J}{-J+\mu_F^2}  \, .
\ee
Of course, the disussion is completely imaterial when considering the Euclidean space theory.

Fixing $\mu_F$ to some observables makes the identification of renormalized buble with its naivelly divergent ancestor  
\be
b_R(q^2;m^2)=b(q^2;m^2) \, . 
\ee

\section{A simple scalar theory in 4d}

Scalar thoeries   are often considered for pedagogical purposes   \cite{RAMON}, \cite{JCCollins},\cite{brown}, \cite{Rupp}, \cite{pokorski}.  
Also here  we begin with the simple model given by  the interacting Lagrangian  
\be   \label{fajfour}
L_{int}=- \frac{h_o}{4!}\phi^4(x) \, ,
\ee
where $\phi$ is the real scalar field. The diagram with a naive log as well the one with  the  quadratic divergence are ilustrated as example.

The bubble diagram (\ref{tchunko})   builds the one loop approximated  scattering amplitude of four $\phi$ particles
 \be 
 M(s,t,u)=h(\mu)+\sum_{q^2=s,t,u}\frac{h^2}{2}b_R(q^2,m^2) \, ,
 \ee
 where $s,t, u$ are  Mandelstam variables composed from the individual incoming and outgoing  particle momenta as usually $s=(p_1+p_2)^2; ...$ and 
 where $b_R(q^2,m^2)$ is $\mu $ dependent renormalized one loop expression for which we have chosen $\ref{treti}$. This implies the renormalized  coupling can be written as
 \be
 h(\mu)=h_0+\frac{h^2}{2}\delta\, .
 \ee
 
Obviously the full amplitude is renormalization invariant $\frac{d M(s,t,u)}{d \mu}=0 $, since  our one loop approximation is equal to the 
original unrenormalized expression
 \be \label{rozptyl}
 M(s,t,u)=h_0+\sum_{q^2=s,t,u}\frac{h^2}{2}b(q^2,m^2) \, . 
\ee

The same equation can be written using the bubble approximation in $FRRS$. Note that the scattering amplitude calculated in FRRS is then locally, but not completely, equivalent to \ref{rozptyl}. 

 In the following, we examine the quadratic divergent loop integral that appears in the self-energy expression of the above theory (\ref{fajfour}). The summed perturbative series (or Schwinger-Dyson equation) for the propagator $G$ is expressed as follows
\bea  \label{prop}
G^{-1}&=&p^2-m^2-\frac{h}{2} a(m^2)-... \, ,
\nn \\
 a(m^2)&=&i\int\frac{d^4k}{(2\pi)^4}\frac{1}{k^2-m^2+\ep} \,. 
\eea
In order to evaluate the tadpole function $a(q)$ one can take  $f=2$ in the Eq. (\ref{tch1}), implying the  second and the first power in the Feynman formula (\ref{fejn}). Actual derivation reads:
\bea 
\nn
a(m^2)&=& -\hat{L}  \int \frac{d^4 k_E}{(2\pi)^4}\int_{\epsilon_z}^1 d z  \frac{ 2 \lambda_F^4 (1-z)}{[(-k_E^2-m^2) z-\lambda_F^2(1-z)+\ep]^3} 
 \\ \nn
&=&\frac{\hat{L}}{(4\pi)^2}\int_{\epsilon_z}^1 d z  \frac{  \lambda_F^4 (1-z)}{z^2[(m^2 -\lambda_F^2)z+\lambda_F^2]}
\nn \\
&\rightarrow& \frac{1}{(4\pi)^2}\left[\lambda_F^2\epsilon_z^{-1}+m^2\left[\ln\frac{m^2}{\mu^2}-\ln [\epsilon_z^{-1} \frac{\lambda_F^2}{\mu^2}\right]\right] \, ,
  \label{intro}
\eea
where again $\mu$ is the renormalization scale and in the second log we have neglected  finite terms against $\epsilon_z^{-1}$.

From now on, we will use the letter $\hat{L}$ not only as a remainder of the limit $\epsilon_z\rightarrow 0$ but in order to stress the entire regularization procedure.
Thus for instance we can   write:
\be 
a(m^2,\lambda_z)=i {\hat{L}} \int \frac{d^4 k}{(2\pi)^4}\frac{1}{ [k^2-m^2+\ep]} =a_R(m)+\delta_a \, ,
\ee
where thre renormalized mass and counterterm parts are    
\bea
a_R(m)&=&\frac{m^2}{(4\pi)^2}\ln\frac{m^2}{\mu^2}+c_a 
\nn \\
\delta_a&=&\frac{1}{(4\pi)^2}\left(\frac{\lambda_F^2}{\epsilon_z}+m^2\ln \epsilon_z+m^2  ln \frac{\mu^2}{\lambda_F^2}\right)-c_a 
\eea
 define number of   renormalization schemes by setting the finite constant $c_a$. 
 For a minimal setting $c_a=0$ the ``renormalized'' propagator $G$ can be written as
\be  
G^{-1}=p^2-m(\mu)^2-\frac{h}{2} a_R(m^2)-... \, ,
\ee
 with renormalized mass $m(\mu)^2=m_o^2+\delta_a$.

This  can be compared with  the dimensional regularization where  conventional result  reads
\be  \label{thoft}
a(m,d)=\frac{m^2}{(4\pi)^2}\left(\frac{2}{\epsilon_d}+ln(4\pi)-\gamma_E+\ln\frac{m^2}{\mu_t^2}\right) \, ,
\ee
where now $\mu_t$ is t'Hooft renormalization scale and  $\epsilon_d=4-d$, $d$  is the number of spacetime dimensions (four in our world).

While the renormalized parts can be chosen identical in two compared schemes above, the main conceptual difference is in presence of   divergent  parts which is proportional to $m$ in DIMR, but is mass independent in our presented scheme. Obviously in our case
\be  
a(0,\lambda_F) \rightarrow \frac{1}{(4\pi)^2}\frac{\lambda_F^2}{\epsilon_z}\, ,
\ee
and masslesness is is not protected either in SRS and  FRRS.  In the later scheme the generated  mass is in fact arbitrary, we get
 \be
 a_{FRRS}= \frac{1}{(4\pi)^2}\mu^2_F .    
\ee

It is well known habit in quantum field theory to accept   $a(0,d)=0$ of the otherwise ambiguous  $o.\infty$ limit in the DIMR and  to prefer the prior of the  massless limit before the spacetime has reached  its correct number of dimensions.

Thus the main, somewhat inconvenient property of the proposed scheme(s)  is that conformal symmetry is not a good symmetry, not only  for massless  $\phi^4$ theory, but quite in general. Masslessness  is    broken  by quantum loops and some tuning is required to bring massless excitation  back into the game. 
To discuss  this critical point further we will start with the calculation of the photon polarization function.  Being able to see, the associated  propagator must describe a massless photon.

For future purposes,  a  more general integrals regularized by the proposed method will be needed. The  list to complete one loop renormalization in the Standard Model reads:
\bea
 I_a(J,\mu_F)=i \int_L\frac{d^4l}{(2\pi)^4}\frac{1}{l^2+J}&=&\frac{1}{(4\pi)^2} \left[\mu_F^2- J \mbox{Lg}(J,\mu_F)\right]   \, ,
\nn \\
 I_b(J,\mu_F)=i \int_L\frac{d^4l}{(2\pi)^4}\frac{1}{(l^2+J)^2}&=&\frac{1}{(4\pi)^2} \mbox{Lg}(J,\mu_F)    \, ,
\nn \\
 I_c(J,\mu_F)=i \int_L\frac{d^4l}{(2\pi)^4}\frac{l^2}{(l^2+J)^2}&=&\frac{1}{(4\pi)^2} \left[\mu_F^2-2 J \mbox{Lg}(J,\mu_F)\right]   \, ,
\nn \\
\label{trelin}
I_d^{\mu\nu}(J,\mu_F)=i \int_L\frac{d^4l}{(2\pi)^4}\frac{l^{\mu}l^{\nu}}{(l^2+J)^2}&=&\frac{g^{\mu\nu}}{4} I_c(J,\mu_F) 
  \, ,
\eea
where the function $J$ depends on the external momenta only and the variable  $\mu_F^2=-\lambda_F^2+\lambda_F^2/(\epsilon_z)^2$
has a dimension of square of the  mass - very meanigfull choice we made for purpose of regularization frrom beginning.    
The symbol  $Lg$ stands for the function \ref{FRS} in FRRS with  the most right hand sides of \ref{trelin} finite in this case.
The same notation is used  for  the difference of two  log functions, i.e.
\be   \label{SRS}
\mbox{Lg}(J;\mu_F)=\ln\frac{-J}{\mu^2}-
\ln\frac{-J+\mu_F^2}{ \mu^2}\rightarrow \ln\frac{-J}{\mu^2}-\delta_c  \, ,
\ee
in SRS where the  limit of small  $\epsilon_z$ is understand. 
If the  first term in (\ref{SRS}) contributes to the renormalized  result and  the second one is sent into the  counterterm, 
then this scheme can be called the minimal SRS  one in analogy to  the case of dimensional renormalization.
The variable  $\mu$ than has a standard  meaning of renormalization scale.

We emphasize again that the finite variable $\mu_F$ that appears in FRRS should not be confused with a renormalization scale.
Rather, it is  the scale  for which values of the momenta the function   $\mbox{Lg}$ begin to vanish  gradually. 

\section{Finite quantum field theory and consequences-  Higgs self mass example}

 As a further  simple example we examine the quark loop contribution to the Higgs selfenergy. This, in adition to Higgs tadpole 
   represent  particularly large correction to the Higgs self-mass  $M-h$ due to the expected size of top-Yukawa coupling $g_Y\simeq 1$ 
 The result, after applying L-operation  reads
\be
\delta M_h(p)=N_c g^2_Y\left[(2 p^2-8 m_t^2)b(p^2,m_t^2,\mu_t^2)-4a(m_h^2,\mu_h^2)\right] \, ,
\ee
where  $N_c=3$ counts number of colors of top quark.

Imposing the following condition:  
\be
  4 N_c g_{Y}^2\lambda_{top}^2-12 h\lambda_{higgs}^2=0 \,.
\ee
 one gets the entire contribution finite in SRS, which however has a no effect on the renormalized $M$, it remains the same.

Quite trivially, there is no infinity left  in FRRS, since all $\mu_i $ in functions $a$ and $b$ are finite but arbitrary.
These new scales  just  need to be adjusted to give the correct Higgs mass. 
 Eventually, other measured observables might help.
There is no  fine tuning in both presented schemes.  Unobservable auxiliar quantities are eaten when compared with observables.  As done in case of Higgs calculation, the  scheme allows to use more scales if needed.
A prize for finiteness in FRRS  is an arbitrarines that need to be fixed. 
Of course , having more options is not always welcome and one can look for princples to reduce them properly. 
The reader may notice that not  only the  philosophy, but also  the asymptotic behaviour of radiative corrections has changed in FRRS. 
  
Before applying in the context of SM and beyond we make a preliminary summary:

-It is evident and will  be exemplified repeatably,  the renormalized parts resulting from diagrams in SRS can be adjusted to align with the established concepts in DIMR, given the latter's relative simplicity in practice.   It is important to note that all Feynman diagrams are inherently finite by their very nature in the FRRS scheme. However, they exhibit distinct asymptotics. 

-The amalgamation of SRS and FRRS in disparate gauge, Higgs, and family (generation) sectors is deemed permissible.

-This approach entails the introduction of more scales in both SRS and FRRS to preserve desired symmetries, such as gauge covariance. 
 Remember the scale origin origin, the scale $\mu_F$  comes from the vanishing limit of $\lambda_F$ , which  originally appears as unit $1=\lambda_F/\lambda_F$ in a given divergent Feynman diagram.
Conformal symmetry is lost in 3+1d from the outset, once more due to the presence of the introduced scales. It is notable that certain simple models of quantum field theory (QFT) exhibit failure in the introduced schemes, with single electron quantum electrodynamics serving as a notable example.

-Emergent phenomena can be described semi-perturbatively. This approach encompasses the phenomenon of chiral symmetry breaking, a process that is known to be responsible for the dynamical generation of the nucleon mass. The subsequent study of this mechanism is reserved for future investigation.  Furthermore, it can account for the gluon mass generation. This approach obviates the necessity for the Schwinger mechanism, which was proposed half a century ago in \cite{jansmit}. This is achieved by acknowledging soft gauge symmetry breaking, a concept that does not pose significant challenges in terms of renormalizability. This phenomenon is illustrated perturbatively in Section \ref{gluonmass}, while a more comprehensive non-perturbative treatment is reserved for a future study.

\section{Photon vacuum polarization  in the SM}  

In this section, the photonic polarization in the Standard Model (SM) will be calculated. It should be noted that, within the class of renormalizable gauges, initially developed for dimensional regularization, the sector of charged vector bosons functions equally well. However, it is important to note that the proposed scheme does not apply to the sector of charged fermions when a single constant scale, $lambda_F$, is chosen.
This particular case will be the focal point of our subsequent analysis. 
The polarization photonic tensor
$\Pi_{\mu\nu}(q)$
 would not satisfy  Ward-Takahashi Idenitity  $q^{\mu}\Pi_{\mu\nu}(q)=0$.
 This failure would result in the photon acquiring a mass. However, it is important to note that restoring gauge invariance is possible in both schemes, i.e., SRS and FRRS.
However, within the SSR, this approach is not natural, as it necessitates distinguishing between leptons and quarks or between generations.
Conversely, the regularized result obtained in FRRS provides a remarkably straightforward solution. This section will emphasize the major related changes and consequences. 
 
 The polarization function is constituted by a loop comprising quark and lepton ($f=q,l$) internal lines.
 The individual contribution is expressed as follows:
 \bea
\Pi^{\mu\nu }_f(q)&=&- i e_f^2 Tr \hat{L} \int \frac{d^4 l}{(2\pi)^4} 
 \gamma^{\mu}\frac{\not l +m}{l^2-m^2+\ep}\gamma^{\nu}\frac{\not l +\not q +m_f}{(l+q)^2-m_f^2+\ep} 
\nn \\
&=&-4 i e_f^2 \hat{L} \int \frac{d^4 l}{(2\pi)^4}\int_0^1 d x  \frac{2 l^{\mu}l^{\nu}-g^{\mu\nu} (l^2-q^2 x(1-x))-2q^{\mu}q^{\nu}x(1-x)+m^2 g^{\mu\nu}}
{[l^2+q^2 x (1-x) -m^2 +\ep]^2} \, ,
 \label{akward}
\eea
where $e_f$ is the charge of the fermion in units of electron charge $e_l=e$; $e_{u,c,t}=2/3e$ $e_{d,s,b}=-1/3$.  

Using  formulas (\ref{trelin}),we can  get the following for  regularized result 
\bea   \label{qed}
\Pi^{\mu\nu  }_f(q)&=&[g^{\mu\nu}q^2-q^{\mu}q^{\nu}]\frac{-4e^2}{(4\pi)^2}\left(\pi(q^2,m,m;\mu_f^2)\right.
\nn \\
&+&\left.\frac{-4e_f^2}{(4\pi)^2} g^{\mu\nu}\mu_f^2\right] \, ,
\eea 
where we have introduced a brief notation for the Feynmann integral over the Lg function
\be  \label{noty}
\pi(q^2,m_1,m_2;\mu_f^2)= \int_0^1 dx 2x(1-x) Lg(q^2 x(1-x)-m_1^2x-m_2(1-x),\mu_f) \, ;
\ee
which is suited for more general case as well.

   Evidently , if the auxiliary parameter $\mu_f$ is choosen not wisely, we can get massive photon in the proposed scheme.
There are more solutions to   avoid this unaccetable  possibility in proposed scheme. Let us start with 
an option that lead to DIMR like result and consider SRS with a $\mu_f $ being a set infinite constants in this case.      

\subsection{Proposition for SRS in the SM }

Upon summing all SM leptons and quarks,  gauge invariance  is recovered by taking 

\be \label{must}
\sum_{F=1}^3(\mu_{l_F}^2+\frac{5}{3}\mu_{q_F}^2)=0 \, 
\ee
where for now $F$ runs over the SM generations.  It is a fine tuned, unnatural  solution in the case of the SM. 
Nevertheless, as an open possibility  we discuss it here. 

It is evident that the complete transversality and masslessness of the photon polarization function can be attained in each family independently, provided that the following adjustments are made: $\mu_{e,\mu,\tau}^2=-5/3\mu_{u,c,t}^2$ and, in the given scheme, the up and down quarks are not distinguished. 
Consequently, the entire renormalized photon propagators in minimal SRS
becomes completely independent on the infinite constant values of $\mu_i=\lambda_i/e_z$. It can be readily demonstrated that the renormalized polarization function is equivalent to that in DIR.

 Of course, a candidate beyond SM that can  provide cancelation of the  second term in \ref{qed}   represent plausible and  natural solution to the problem as well.
Remarkably, the associated sum rules count the group factors independently of masses of particles in virtual loops.

In closing this section, the observation is made that the charged SM vector bosons are of limited utility in this regard. It is important to note that the condition necessary to cancel the longitudinal terms, proportional to the inverse of the regulator, $\epsilon_z$ reads
\be
3(\lambda_W^2-\lambda_T^2)- 4\sum_F^{N_F}(\frac{5}{3}\lambda_{qF}^2+\lambda_{lF}^2)=0
\ee 
where $\lambda_W$ is the auxiliar parameter in the loop with ghosts, Goldsteones and two W-propagators, $ \lambda_T $ appears in the regulated expression for the tadpole, subscripts $q$ and $l$ stands for universal scale for  quarks and  leptons,  $N_F=3$ is the known number  of SM families. 
Permitting greater autonomy in the auxiliary, non-observable parameter space of SRS regulators yields equivalent results to the established SM expressions in DIMR. 

\subsection{Proposition for FRRS}
When using the finite constant $m_f$, we wouldn't be able to select a detailed, concrete scheme and express it in FRRS because there aren't any established principles. However, a wide freedom in the selection of parameter $\mu_f$ offers an elegant
solution. Let's consider the following:

\be \label{choice}
\lambda_f/\epsilon_z\rightarrow \mu_f^2=-q^2 
\ee
where $q$ is the external momentum. i.e. the momentum of the photon in our case.

In this case , the masslessness of photon is not sacrified. This  choice  is equivalent to tiny shifting of the residue of the photon propagator by the following   amount  
\be 
C=\sum_f\frac{4e_f^2}{(4\pi)^2}
\ee
at one loop. It is not an observable as it can be absorbed by finite renormalization.

 The inspection of the gauge propagator can also help identify the finite redefinition of the gauge-fixing parameter, $\xi$. The calculation shows
 \be  \label{foton}
 G^{\mu\nu}(q)=G_T(q^2)[-g^{\mu\nu}+\frac{q^{\mu}q^{\nu}}{q^2}]+\xi^{new}\frac{q^{\mu}q^{\nu}}{{q^2}^2}
 \ee
 in the FRRS, where the transverse propagator function  (we have suppressed the color indices) reads
 \bea
 G_T(q^2)&=&\frac{1}{q^2(1+\Pi_T(q^2)-C)} 
 \nn 
\\
\label{shift}
\xi_{new}&=&\frac{1}{\xi^{-1}+C} \, .
 \eea

One loop prediction of FRRS is that there will be an absence of Landau poles at deep ultraviolet and a change in the running of QED effective charge at larger $q^2$. Considering these changes, we have arrived at a new expression for the polarization function
\be
\Pi_T(q)=\sum_f \frac{-4e_f^2}{(4\pi)^2} \pi(q^2,m_f,m_f;-q^2)
\ee
with the  $\pi(..)$ i introduced by (\ref{noty}).  

 If the FRRS is in the description of Nature, we have substantial changes in the  Grunberg's effective charge $\alpha_{QED}(q^2)$, defined as usualy:
 \be
 \alpha_{QED}(q^2)=\frac{\alpha(0)}{1-\Pi_T(q^2)}
 \ee 
  in both spacelike and timelike scales, noting that $\alpha_{QED}(0)=\alpha(0)=1.137$ being atomatically satisfied in FRRS
 since quite obviously $\Pi_T(0)=\pi(0,m,m,0)=0$.
  The proposed scheme is not completely equivalent to other known renormalization schemes, the calculated running charge differs at deep ultraviolet.  Hence quantitative  changes associated with proposed sheme can be traced or excluded via experiments. 
   
 Notably, the effective charge does not grow to infinity, it does not posses a Landau pole at some transPlankian scale. The universal bare charge
 of SME fermions are finite quantities not difficcult  to predict within proposed perturbative FRRS:
 \be
 \alpha_0=\frac{\alpha(0)}{1-C}
 \ee
 where $C$ collects finite small  pieces from primitively divergent loop corrections.

\section{NonAbelian theory, Gluon polarization function and QCD effective charge }  \label{gluonmass}

In the previous section, we presented two different new methods for calculating radiative corrections to the photon propagator.
Although one is a limiting case of the other, and although both preserve gauge invariance, they lead to inequivalent results at one loop approximation. 
One makes the theory finite, i.e. it leaves bare charge finite, the second scheme can be adjusted to be equivalent to DIMR, at least it is true for one loop.
In both cases there is no room for improvement by renorgroup equations. The unrenormalized Greens functions are individually manifestly independent of the renormalization scale.     
The implementation of the proposed schemes in the context of massless Yang-Mills theory and consequently in perturbative quantum chromodynamics (pQCD) follows the same paths. An identification of SRS and DIMR is exemplified, although it requires a very unnatural adjustment of the counterterms.  This follows from the fact that when resembling DIMR one needs constant infinite regulators and require their sum to vanish to ensure gauge invariant the massless limit.  To meet this challenge, the associated calculation bears a resemblance to photonic vacuum polarization, there are just more possibilities since there are more diagrams in the sum.

However, the focus now shifts to the second and more natural option, which is the FRRS. To address this challenge, we first calculate the gluon polarization.   To compare the quantum consequences that arise in finite QCD, we present the result for the QCD effective charge in FRRS QCD. Making all quantum corrections finite leads to the evolution of the color charge: according to asymptotic freedom it does indeed become smaller at large $q^2$, but it does not vanish at ultraviolet, becoming frozen and asymptotically flat above the electroweak scale $|q^2|>200GeV$.    We do not discuss the non-perturbative issue of the gluon mass generation observed in the lattice evaluation, which phenomenon has also raised controversies regarding its origin in the Schwinger pole structure of the gluonic vertices \cite{MAME2019,MAVU2022,Ag2023}. 
 We also do not discuss the issue of confinement and present the perturbative gauge invariant one-loop results in FRRS for the massless gluon.
 
The subsequent discussion aims to elucidate the necessary conditions for a symmetry preserving treatment finite renormalization scheme. 
Since the results can be derived very straightforwardly from the regularized but not yet renormalized expansion, we discuss the individual schemes at the end. The first steps are to deal with primitively divergent integrals of one-loop QCD, their form can be found in various textbooks. We work in the Feynman gauge for simplicity.
The standard calculation includes loops with ghosts, in which case we label the regulator as $\lambda_F=\Lambda_c$, analogously the loop with two gluonic internal lines will use $\lambda_g$, a tadpole with a single gluon line will use $\lambda_t$, and when the full QCD case is considered, the loop with quark lines will be treated with the $\lambda_q$ regulator.
 
In the  Feynman gauge, the contribution for the loop with two internal gluonic lines is expressed as follows:
\be 
\Pi_g^{\mu\nu ab}(q)=\frac{g^2}{2}i{\hat{L}}   \int \frac{d^4 l}{(2\pi)^4} \frac{ T(A) \delta^{ab}  N_{\mu\nu}(q,l)}{[l^2+\ep][(l+q)^2+\ep]}\, ,
\ee
where 
\be
N^{\mu\nu}(q,l)= 10 l^{\mu}l^{\nu}+5(l^{\mu}q^{\nu}+l^{\nu}q^{\mu})-2 q^{\mu}q^{\nu}+g^{\mu\nu}( 2l^2+2l.q+5q^2)\, ,
\ee
where $T(A)=3$ for o $SU(3)$.

% structure constant $f^{cad}f^{cbd}=T(A)\delta^{ab}$.
  
Proceeding further one gets
\bea 
\Pi_g^{\mu\nu ab}(q)&=&\frac{g^2}{2} T(A) i\delta^{ab} {\hat{L}}\int_0^1 d x \int \frac{d^4 l}{(2\pi)^4}\left[\frac{9}{2}\frac{g^{\mu\nu} l^2}{[l^2+q^2 x(1-x)+\ep]^2}\right.
\nn \\
&+&\left.\frac{g^{\mu\nu} q^2(2x^2-2x+5)+q^{\mu}q^{\nu}}{[l^2+q^2 x(1-x)+\ep]^2}\right]
\eea

Regularizing by $L$-operation one gets
\bea 
\Pi_g^{\mu\nu ab}(q)&=&\frac{g^2}{2(4\pi)^2} T(A) \delta^{ab}  
\int_0^1 d x 
\left(\frac{9}{2}g^{\mu\nu} [\lambda_g^2/\epsilon_z-q^2 x(1-x)Lg(q^2x(1-x),\mu_g^2)]\right. 
\nn \\
&+&\left.\frac{g^{\mu\nu}}{2} q^2(2x^2-2x+5)+\frac{q^{\mu}q^{\nu}}{2}(10x^2-10x-2)
Lg(q^2x(1-x),\mu_g)\right)
\eea

Regularized gluonic tadpole contributes by the following amount:
\be
\Pi_t^{\mu\nu ab}=\frac{g^2}{(4\pi)^2} 3 T(A) \delta^{ab} g^{\mu\nu} (\frac{\lambda_t^2}{\epsilon_z}-\lambda_t^2) \, .  
\ee 
 
Ghost loop gives  contribution to metric tensor as well as to transverse part:
\bea 
\Pi_c^{\mu\nu ab}&=&(-1)\frac{g^2}{(4\pi)^2} T(A) \delta^{ab}  \int_0^1 d x  
\left[\frac{-g^{\mu\nu}}{2} [\lambda_c^2/\epsilon_z-q^2 x(1-x)Lg(q^2x(1-x),\mu_c)]\right.
\nn \\
&+&\left[q^{\mu}q^{\nu}(1-x)x Lg(q^2x(1-x),\mu_c^2)\right. \, .
\eea

The meaning of the introduced functions and scales, i.e. the relation between different $\mu_i$ and $\lambda_i$, is the same as in the previous sections.

In adition we take  $\lambda_g=\lambda_c$ for simplicity. By summing the three terms, we complete the gluon polarization function in pure Yang-Mills theory. To preserve transversity, masslessness and to get a DIMR, the necessary condition is
\be  \label{jang}
3\lambda_t^2+4\lambda_g^2=0\, .
\ee 

 Thus, to show equivalence with DIM for pure gauge massless Yang-Mills theory, one has to take some regulators negative.
Although the signs of $\lambda_i$ are irrelevant for the renormalized gauge, and although famous DIMR results are reproduced for the renormalized polarization function, the loop-dependent choice of $\lambda_i$ and the associated manipulation necessary to restore gauge invariance is certainly the most awkward feature of the presented SRS scheme.

Before presenting the results for FRRS we  complete the gluon polarization function in QCD and  present the perturbative contribution due to quark loop.
 Up to the group factors it has the identical form known from QED:
\bea
\Pi^{\mu\nu ab}_q(q)&=&- i g^2 T(R) \delta^{ab}Tr \int \frac{d^4 l}{(2\pi)^4}  \gamma^{\mu}\frac{\not l +m}{l^2-m^2+\ep}\gamma^{\nu}\frac{\not l +\not q +m}{(l+q)^2-m^2+\ep} 
\nn \\
&=&-4T(R) ig^2 \delta^{ab} \int \frac{d^4 l}{(2\pi)^4}\int_0^1 d x  \frac{2 l^{\mu}l^{\nu}-g^{\mu\nu} (l^2-q^2 x(1-x))-2q^{\mu}q^{\nu}x(1-x)+m^2 g^{\mu\nu}}
{[l^2+q^2 x (1-x) -m^2 +\ep]^2} \, ,
 \label{vstup}
\eea
where $T(R)=1/2$ for quarks in fundamental representatuion of $SU(3)$ group.
Using  formulas (\ref{trelin}) the regularized result reads 
\bea \label{lopata}
\Pi^{\mu\nu ab }_q(q)&=&[g^{\mu\nu}q^2-q^{\mu}q^{\nu}]\frac{-4g^2}{(4\pi)^2} T(R)\delta^{ab}\pi(q^2,m_q,m_q;\mu_F^2)
\nn \\
&+&\left.\frac{-4g^2}{(4\pi)^2} T(R)\delta^{ab}g^{\mu\nu}(\frac{\lambda_q}{\epsilon_z}-\lambda_q)\right] \, ,
\eea

Summing all terms together one gets for the polarization function with $n_q$ quarks
\bea  \label{barbara}
\Pi^{\mu\nu ab}(q)&=&\frac{g^2}{(4\pi)^2}  \delta^{ab} \frac{g^{\mu\nu}}{\epsilon_z}
\left(3T(A)\lambda_t^2 +4T(A)\lambda_g^2- 4 n _q T(R)\lambda_q^2\right)
+ \left(g^{\mu\nu}q^2-q^{\mu}q^{\nu}\right)\delta^{ab}\Pi(q^2)
\nn \\
\Pi(q^2)&=& 
\frac{g^2}{(4\pi)^2}
\left(T(A) \frac{10}{6} \ln\frac{-q^2-\ep}{\mu^2}
-8T(F)\sum_q \int dx x (1-x) \ln \frac{-q^2x(1-x)+m_q^2-\ep}{\mu^2} 
+K\right) \, ,
\eea
where just for puprose of SSR, $K$ stands for the collection of unimportant  constant terms including log of regulators as well (here we take $\lambda_c=\lambda_g$  for simplicity).

Requiring the gluon mass to vanish $m_g^2=0$, were one loop expresion for the gluon mass  is 
\be \label{ojoj}
m_g^2=\frac{g^2}{(4\pi)^2}  
\left(3T(A)\mu_t^2 +4T(A)\mu_g^2- 2 n _q T(R)\mu_q^2\right)
\ee
one gets the renormalized  polarization function in minimal SRS  exactly  identical to its counterpartner calculated in 
DIMR. 

\subsection{FRRS in QCD}

 Having proved the one-loop equivalence of SSR and DIMR, a non-trivial question is the application of FRRS to QCD.
In the QCD partonic model, accounting for the vertex correction at a loop then, counting the leading logs (Lg functions in our case) then one can define the effective QCD charges on the basis of studying a given partonic scattering process.  
Recall here that the standard expected behavior $\alpha_{S} \rightarrow 0 $ for asymptotically large momenta $Q^2$ is an inherent consequence of known textbook renormalization programs.  The asymptotic freedom in its extreme form of an asymptotically zero gauge coupling can be traced as a consequence of subtracting infinities during the renormalization program.   Can we get a known process-independent QCD running charge in FRRS? An immediate answer is NOT, if one wants to keep asymptotic freedom in its extreme version $\alpha_{S} \rightarrow 0 $. 
Assuming that the effective charge can be defined in our case, then its evolution, governed by the presence of $Lg$ functions, slows down for $q^2>>\mu_{max}^2; \mu_{max}=max{\mu_i} $ and the coupling becomes flat. In other words, due to the finiteness of the radiative corrections, we get a supersymmetric scenario without supersymmetry being imposed.

The freezing of the coupling at some high $Q^2$ is a prediction of the FRRS, and to make a first comparison we calculate the QCD effective charge.
To do this, we consider light quark scattering at a loop and sum or loop leading $Lg$ terms that appear in such a process. As usual for partonic calculation,
 we ignore magnetic moment couplings, ignore the confinement problem, and assume on-shell light (u or d) quarks in the initial and final states.  
Then the high energy scattering is governed by a single (loop dressed) gluon exchange, with the line of the gluon propagator connected to the (loop dressed) quark gluon vertex. 

Based on the approximated scattering amplitude (spinors are omitted) of the $qq\rightarrow qq$ process, the effective charge is defined as
\be
\alpha(q^2)_{S}\gamma^{\mu}\times \gamma^{\nu}\frac{ g^{\mu\nu}}{q^2}
 \ee
where all  longitudinal pieces of the gluon propagator has been ignored, since the cancel against on shell quarks and 
the effective charge reads
\be
\alpha_{S}(q^2)=\frac{g^2}{(4\pi)[1-\Pi_T(q^2)]}[1+V_A(q)+V_{NA}(q)]
\ee
In addition to the gluon polarization function, which characterizes the gauge boson propagator, one has to consider on the loop $Lg$ the corrections $V_A$ and $V_{NA}$ from the abelian and non-abelian one-loop diagrams, which contribute to the $\gamma_{\mu} T^{a,b}$ component of the quark-gluon vertex.
As follows from the derivation, one can get vertices in FRRS from the known DIMR expression by replacing identified DIMR protocols by $Lg$ functions in expressions for the vertices, i.e. DIMR can be used immediately for this purpose. Here is the result
 \be
 V_A(q)+V_{NA}(q)=(C(R)+T(A))\frac{2g^2}{(4\pi)^2}\pi(q^2,m,m,-q^2)
 \ee
 where $C(R)=\frac{N^2-1}{2N}=4/3$ and  the argumet of the functions $\pi$ means taht now we take  $\mu_F^2=-q^2$ in the expression for the $Lg$. 
% srednicky 442

Indeed, to get the finite gluon polarization function $\Pi_T$ satisfying atomatically the one-shell condition
$\Pi_T(0)=0$ one needs to take $\mu_i^2=-q^2$
in all $i$ subdiagrams discussed above.   Hence, to take $\mu_F^2=-q^2$ in the expressios for the quark gluon vertex 
turns to be quite natural choice.

In FRRS the  gluon propagator then  reads
\be
\delta_{AB}G^{\mu\nu}(q)
\ee
where $G^{\mu\nu}(q)$ has the form identical to photon propagator \ref{foton}.
The polarization function reads in loop
\be
\Pi_T(q^2)= 
\frac{g^2}{(4\pi)^2}
[5 T(A) \pi(q^2,0,0,-q^2)-8T(R)\sum_q  \pi(q^2,m_q,m_q,-q^2)]
\ee
where again the sum runs over the quark flawour.

To gain insight into what the changes are from the traditional scenario, we evaluate the FRRS running coupling by matching its value to that determined from the experiment $\alpha(M_Z)=0.1178$, which is a 2023 world average \cite{PDG} or $\alpha(M_Z)=0.1157$ \cite{exper} as determined from the H1 experiment alone. Here we simply assume that the associated scheme changes are small at this large scale. The resulting effective charge is shown in the figure \ref{alfa}. The gluon form factor \ref{gfor} is shown for both the spacelike and the time-like regions of the square of the momenta.

Several striking features are observed in the calculated effective charge.

\begin{itemize}

\item It can be continued smoothly down to low $q^2$, although the confinement phenomena and the non-perturbative corrections can change running below a few $GeV$.  

\item There is no Landau pole at $\Lambda_{QCD}$, the perturbatively calculated FRRS coupling behaves smoothly at low energies. 

\item At very low $q^2$ stays bellow critical strength needed for the correct chiral symmetry breaking phenomenon.
      Other contributions, likely other components of the quark-gluon vertex seems to be
       important in the Feynman gauge in presented scheme.   
 
 \item At few GeV, the couplings are identical or slightly weaker then expected values from quarkonium physics. 
  
 \item Bare coupling can be determined, which is $\frac{g^2}{(4\pi)^2}=0.75 $ in our case. The effective charge freezes in the infrared as well as in the ultraviolet. The latter predicts $\alpha_S(\infty)\simeq 0.09 $ in  loop approximation for six quarks observed so far in Nature.
 
 \end{itemize}

 Last but not least, one could mention that the auxiliary gauge fixing paaremetr has been lowered by the identity \ref{shift} by the amount determined by $C_{qcd}$.
\be
 C=\frac{g^2}{(4\pi)^2}  [3T(A) +4T(A)- 12 T(R)]=0.75 \, .
\ee
In other words, a selfconssistent generation of constant (unphyssical?) renormalizable longitudinal part is associated  phenomena 
in FRRS. We started with the Feynman gauge ( $\xi=1$ in our notation) , but ended with $\xi\simeq 4/7$, chalenging thus academical problem with self-consistence, assuming the  gauge term plays no  more than its auxiliar role in physics.

\begin{figure}[htb]
\centerline{\includegraphics[width=10.0cm]{alfastrong.eps}}
\caption{\label{alfa}Effective charge in finite QCD }
\end{figure}
 
\begin{figure}[htb]
\centerline{\includegraphics[width=10.0cm]{alfy.eps}}
\caption{\label{gfor} Gluon form factor in the timelike and spacelike range of momenta. }
\end{figure}

\section{ABJ  triangle anomaly} 

 In general, quantum anomalies are quantum corrections that do not respect the Ward identities in any known renormalization scheme. In addition, anomalous terms exist.
 Among them, a chiral anomaly \cite{ABJ1},\cite{ABJ2} played the key historical role. 
The Standard Model is an anomaly-free theory, chiral anomaly cancellation between leptons and quarks diagram for each family individually, and thus a sensitivity to a given RS schemes is not only of urgent need. Nevertheless, it is interesting to see how the presented renormalization schemes agree with the evaluation of individually divergent fermion loop diagrams. Compared to the vacuum self-energy diagrams studied in the previous section, the naive degree of divergence is reduced. It entails that the final result is not uniquely defined due to the repeated appearance of the problematic double limit $0.\infty$ in SRS.  Both Ward identities in the game, the electromagnetic and the chiral one, are satisfied in the presented renormalization scheme only if we accept the preferred ordering of the aforementioned double limit.

Considering single quark (or lepton), the chiral anomaly would appear in the well known sum of  triangle diagrams 
which would violate at least one of the Ward identity listed here: 
\be \label{WTI}
- (p+q)^{\mu}\Gamma^5_{\mu\nu\delta}(p,q,m)= 2m \Gamma^5_{\nu\delta}(p,q,m)\, \, \, ;\,\,\, p^{\delta}\Gamma^5_{\mu\nu\delta}(p,q,m)=p^{\nu}\Gamma^5_{\mu\nu\delta}(p,q,m)=0 \, ;
\ee 
 where $m$ is the  quark(lepton) mass, which appears  in all three quark propagators inside the triangle loop and we will  adopt standard  conventions used in the textbook  \cite{pokorski},
  i.e. $p$ and $k$ are outgoing photons, implying that $p+q$ is the four-momentum associated with external line of axial-vector vertex (i.e. $Z^0$  boson corresponds with the external line in case of the Standard Model).
    For the calculation using the Pauli-Villars regularization, we refer to the standard textbook \cite{pokorski}.
 For subtleties associated with the use of DIMR when dealing with intrinsically 4-dimensional objects such as the $\gamma_5$ matrix or the Levicivita tensor $\epsilon_{\alpha\beta\nu\mu}$, see for example \cite{NOVOTNY}.

It is instructive to  consider the electromagnetic WTI using only  a single fermion loop 
\bea \label{had}
 p^{\delta}\Gamma^5_{\mu\nu\delta}(p,q;m)&=& 2 i Tr \hat{L} \int \frac{d^4 k}{(2\pi)^4}  \frac{\not k+\not q +m}{(k+q)^2-m^2+\ep}\gamma^{\nu}\frac{\not k - \not p  +m}{(k-p)^2-m^2+\ep}\gamma^{\mu}\gamma_5 
\nn \\
&-&2 i Tr \hat{L} \int \frac{d^4 k}{(2\pi)^4}  \frac{\not k+\not q +m}{(k+q)^2-m^2+\ep}\gamma^{\nu}\frac{\not k  +m}{k^2-m^2+\ep}\gamma^{\mu}\gamma_5 
\nn \\
&=& -i \hat{L} \int \frac{d^4 k}{(2\pi)^4}  \frac{8 i\epsilon_{\rho\nu\sigma\mu}( q^{\rho} p^{\sigma}+ k^{\rho} p^{\sigma}- q^{\rho} k^{\sigma})}{[(k-p)^2-m^2+\ep][(k+q)^2-m^2+\ep]} \, , 
\eea
where again the symbol ${\hat{L}}$ means that we should perform (or finish) the  $L$-operation.  Note 
the second line in Eq. (\ref{had})  vanishes exactly, since being proportional to   $q^{\alpha}q^{\beta}\epsilon_{\alpha\beta\nu\mu} I(q,m)$ where $I(q,m)$ is the finite  function.

In the result (\ref{had})  we will use the Feynman variable $x$ in order to match the denominators and  the variable $z$ for purpose of $L$-operation.
It gives us
\be
 - i \hat{L} \int \frac{d^4 k}{(2\pi)^4} \int_{\epsilon_z}^1 d z\int_0^1 dx
i 8 \Gamma(3) z  \lambda_F^2 \epsilon_{\rho\nu\sigma\mu}
\frac{[q^{\rho} p^{\sigma}+ k^{\rho} p^{\sigma}- q^{\rho} k^{\sigma}]}{z^3[k^2+k.q x -k.p(1-x)+q^2x+p^2(1-x)-m^2-\lambda_F^2\frac{1-z}{z}+\ep]^3}  \, .
\ee
 After changing the ordering of  integrations one gets finite momentum  integral  and we perform
the shift of the momenta by making  the substitution $k=k_{new}-q x+p (1-x)$ and integrate over the momentum $k_{new}$. The result can be written as follows
\be
-\hat{L}\frac{8i}{(4\pi)^2} \int_{\epsilon_z}^1 d z\int_0^1 dx \frac{\lambda_F^2 \epsilon_{\rho\nu\sigma\mu}\left(q^{\rho} p^{\sigma}+ x q^{\rho} p^{\sigma}- q^{\rho} p^{\sigma}(1-x)\right)}
{z\left[\left((p+q)^2 x(1-x)-m^2\right)z-\lambda_F^2(1-z)+\ep\right]} \, ,
\ee 
which in the limit $\epsilon_z \rightarrow 0 $ provides the following ambiguous result: 
\bea \label{emgw}
-p^{\delta}\Gamma^5_{\mu\nu\delta}(p,q;m)&=&
-\frac{8i}{(4\pi)^2} \epsilon_{\rho\nu\sigma\mu}\left[q^{\rho} p^{\sigma}- q^{\rho} p^{\sigma}\right]
\int_0^1 dx (1-x)\left[\ln{\frac{(q+p)^2x(1-x)-m^2+\ep}{-\mu^2}}-\ln{\frac{\lambda_F\epsilon_z^{-1}}{\mu^2}}\right]
\nn \\
&=&[0].[ \infty] \, .
\eea 

Not well defined expression (\ref{emgw}) is how anomalous diagrams are reflected in presented SRS.
Needless to say, that considering the sum of the lepton and the quark loops, one gets the analogue of the 
second line in the Eq. \ref{emgw} but  terms which include $\epsilon_z$ vanishes mutually providing thus very correct maintaining of  the limit $\epsilon\rightarrow 0$. 
Then the WTI $ -p^{\delta}\Gamma^5_{\mu\nu\delta}(p,q;m)=0$ is valid without any ambiguities in the Standard Model.

The treatment of  the chiral Ward identity follows very similar steps.
 Here we refer to the page 462 Eq. (13.15) in the textbook \cite{pokorski}, which we start with, however without presence of  Pauli-Villars regulators , but rather with the symbol 
 of $L$-operation instead.  It reads
\bea \label{axial}
 -(p+q)^{\mu}\Gamma^5_{\mu\nu\delta}(p,q;m)&=& 2 i\hat{L} Tr \int \frac{d^4 k}{(2\pi)^4}  \frac{\not k+\not q +m}{(k+q)^2-m^2+\ep}\gamma^{\nu}
 \frac{\not k +m}{k^2-m^2+\ep}\gamma^{\delta}\gamma_5 
\nn \\
&-&2 i \hat{L} Tr \int \frac{d^4 k}{(2\pi)^4} \gamma^{\nu} \frac{\not k +m}{k^2-m^2+\ep}\gamma^{\delta}\frac{\not k-\not p  +m}{(k-p)^2-m^2+\ep}\gamma_5 
\nn \\
&+& 2m \Gamma^5_{\alpha\beta}(p,q) \, .
\eea

Performing  $L-$operations one gets zero separately for each of the  first two lines in the following form 
\bea
\epsilon_{\nu\delta\alpha\beta}p^{\alpha}p^{\beta} b(p,m)&=&0.\infty \, ,
\nn \\
\epsilon_{\nu\delta\alpha\beta}q^{\alpha}q^{\beta} b(q,m)&=&0.\infty  \, ,
\eea   
where again  $ b(q,m)$ stands for the  function (\ref{treti}).

The third line in the Eq. (\ref{axial})  is proportional to the pseudoscalar-vector-vector triangle i.e. to the  rhs. of axial WTI and finishes formally the derivation.
Just for completeness, it reads
\be \label{piondecay}
 \Gamma^5_{\mu\nu}(p,q)= 2 i Tr \int \frac{d^4 k}{(2\pi)^4}  \frac{\not k+\not q +m}{(k+q)^2-m^2+\ep}\gamma^{\nu}
 \frac{\not k +m}{k^2-m^2+\ep}\gamma^{\delta}  \frac{\not k - \not p  +m}{(k-p)^2-m^2+\ep}\gamma^{5} 
\ee
and since being finite, this term is independent of the renormalization scale $\mu$ as well as of  the regulator $\epsilon_z$.

The get the result of the calculation  in the limiting FRRS is more then  straigforward, kepping the triangles  finite
 then EMG WTI is uniquely satisfied $-p^{\delta}\Gamma^5_{\mu\nu\delta}(p,q;m)=0$ as well as the AWTI does,
since
\be
\epsilon_{\nu\delta\alpha\beta}p^{\alpha}p^{\beta} b_{FRRS}(p,m)=0 \, .
\ee

There is no anomaly presented even if the $SM$ generation would be incomplete ( say there are more $SU(2)$ quark dublets and singlets then leptonic ones.

 \section{Overlapping divergences- sunset diagram }
\label{sunrise}

The term "diagrams with overlapping divergences" is employed to denote those diagrams in which multiple sub-loops share the same propagator line(s), and at least two separate momentum integrations are divergent.
A meaningful RS could provide a result for multiloop Feynman diagrams with overlapping divergences as well. 
 In a manner analogous to the context of dimensional renormalization, where the application of dimensional regularization is accompanied by the necessity for meticulous removal of infinities, a similar level of rigor is demanded in  presented schemes.
In this instance, the repeated use of the $L$-operation results in a "remaining singularity" that is transferred to additional Feynman variable integrals. This "remaining singularity" must be subtracted.
 As is customary, such infinities can be subtracted algebraically and removed by introducing a subtraction polynomial with a finite number of terms. The renormalization program is then finished by identifying the polynomial
  coefficients with counter-term part of  the Lagrangian. Here we will show the appropriate treatment  for the case of sunset diagram.

We do not discuss FRRS here, due to the existing ambiguities that have yet to be resolved.

The sunset diagram is the   two loop irreducible diagram in $\Phi^4$ 
that contributes to the scalar field selfenergy and there are certainly more  possibilities how it can be renormalized in the proposed SRS. 
 Here we present a simple  method, which is particularly suited for   generic case of multiloop diagrams in  the Standard Model. 
 The method is composed of two steps. In the first step, the integrations over the momenta are performed by repeated use of $L$-operation, and in the second step, the subtraction of infinities is made algebraically.

Two limits  (\ref{tch3}) are required, which is now what the symbol of $L-$operation stands in case of  the following expression for the sunset diagram
\be
Sun(q^2,m^2)=i \hat{L} \int \frac{d^4 k}{(2\pi)^4}  i\int \frac{d^4 l}{(2\pi)^4}   \frac{1}{[k^2-m^2+\ep][(k+q+l)^2-m^2+\ep][l^2-m^2+\ep]}\, , 
\ee
where we have skipped a constant symmetry  prefactor (which is $h^2/6$ if  $\Phi^4$ theory is considered). 

 Performing the first  L-operation for purpose of the  integration over the  
 momentum $k$  we should  get the  following expression
\be 
Sun(q^2,m^2)=i \hat{L} \int \frac{d^4 l}{(2\pi)^4}\frac{b((q+l)^2,m^2)}{[l^2-m^2+\ep] }\, ,
\ee
where $b$ stand for  the bubble integral (\ref{treti}).

Let us  perform the substitution $x=(1-z)/2$ 
in the integral expression for  $b((q+l)^2)$ , which after a short algebra gives 
\be 
Sun(q^2,m^2)=ln(1/4) \frac{a(m^2)}{(4\pi^2)}+ \delta_b a(m^2)+
 i\frac{\hat{L}}{(4\pi)^2} \int \frac{d^4 l}{(2\pi)^4}\int_0^1 dz
 \ln\frac{(q+l)^2  (1-z^2)-4m^2+\ep}{-\mu^2}\frac{1}{l^2-m^2+\ep} \, .
\ee
After per partes integration we can get
\be 
 Sun(q^2,m^2)=K_{8.4} 
+i\frac{\hat{L}}{(4\pi^2)} \int \frac{d^4 l}{(2\pi)^4}\int_0^1 dz
\frac{2 z^2 (q+l)^2 }{(1-z^2) [(q+l)^2-\frac{4m^2}{1-z^2}+\ep][l^2-m^2+\ep]} \, ,
\ee
%page -2- Dolni Zleb
%page 2 silvestr
where we have introduced shorthand notation  for the constant 
$$K_{8.4}=\ln(\frac{m^2}{\mu^2})\frac{a(m^2)}{(4\pi^2)}+\delta_b a(m^2)$$  .

Let us perform substitution $z\rightarrow \omega$ such that  
\be  \label{subst}
\omega=\frac{4m^2}{1-z^2} \, ,
\ee
 which allows us write the expression for  sunset in terms of   single loop bubble and tadpole functions $b$ and $a$ respectively
\bea \label{result2}
 Sun(q^2,m^2)&=&K_{8.4}
+i \frac{\hat{L}}{(4\pi^2)}\int \frac{d^4 l}{(2\pi)^4}\int_{4m^2}^{\infty} d\om \frac{\sqrt{1-\frac{4m^2}{\om}}}{\om}
\frac{ (q+l)^2 }{[(q+l)^2-\om+\ep][l^2-m^2+\ep]} \, .
\nn \\
&=&K_{8.4}+\frac{1}{(4\pi^2)} \int_{4m^2}^{\infty} d\om \sqrt{1-\frac{4m^2}{\om}}\left[\frac{a(m^2)}{\omega}+\delta_b\right]
\nn \\
&+&\frac{1}{(4\pi)^2} \int_{4m^2}^{\infty} d\om \sqrt{1-\frac{4m^2}{\om}} b(q^2;\omega,m^2) \, ,
\eea
 where  the   function $b(p^2;\omega,m^2)$ is the  expression for the bubble integral with 
different masses: $\sqrt{\omega}$ and $m$. It is worthwhile to calculate  this integral  separately here 
and the $L$ regularization gives as
\bea  \label{blob2}
b(p^2;\omega,m^2)&=&i \hat{L} \int\frac{d^4l}{(4\pi)^4}\frac{-\lambda_F^2}{[(p+l)^2-\omega+\ep][l^2-m^2+\ep][-\lambda_F^2]}
\nn \\ 
%&=&i \hat{L}  \int\frac{d^4l}{(4\pi)^4}\int_{\epsilon_z}^1 d z \int_0^1 d y
%\frac{-2 \lambda_F^2 z }{z^3[l^2+p^2 y (1-y)-m^2 y-\omega(1-y)-\lambda_F^2\frac{1-z}{z}]^3}
&=&\frac{1}{(4\pi)^2}\int _0^1 dy \ln\frac{p^2 y(1-y)-m^2 y-\omega(1-y)+\ep}{-\mu^2}\, +\delta_b \, ,
\eea
where $\delta_b $ is already defined in the Eq. (\ref{treti}).

Obviously, the multiple applied L-operation does not sent all divergences into the regulator dependent part completely.
 there are further linearly as well as logarithmically  divergent terms in the $\omega$ variable.
These terms, together with constant terms, can be subtracted algebraically by and sent to counterterms.
For this purpose we substitute (\ref{blob2}) into  the expression  (\ref{result2}) 
and integrate  per-partes with respect to the variable $y$ :
\bea \label{result3}
 Sun(q^2,m^2)&=&\frac{a(m^2)}{(4\pi)^2}\ln(\frac{m^2}{\mu^2})+\delta_b a(m^2)+
 \frac{1}{(4\pi)^2}   \int_{4m^2}^{\infty} d\om \sqrt{1-\frac{4m^2}{\om}}\left[\frac{a(m^2)}{\om}
+\frac{1}{(4\pi)^2} \ln\frac{\om}{\mu^2}+\delta_b\right] 
\nn \\
&-&\int_{4m^2}^{\infty} d\om \int_0^1 \frac{d y}{1-y} \frac{\sqrt{1-\frac{4m^2}{\om}}}{(4\pi)^2} 
\frac{q^2(1-2y)-\om+m^2}{q^2-\Omega+\ep} \, ,
\eea
where we have labeled 
\be \label{subst2}
\Omega=\frac{m^2 y+\omega(1-y)}{y(1-y)}.
\ee

Now it is  obvious that   the Eq. (\ref{result3}) takes the form of  un-subtracted  dispersion relation, which means that the 
infinities  can be absorbed into the mass counter-term 
\be \label{firstsubs}
\delta m= Sun(p^2=\zeta,m^2) +c_1 \, ,
\ee
and  by the field redefinition 
\be
Z=1+\delta_{\phi}\,\,\,,\,\,\,\, \delta_{\Phi}=\frac{d Sun(x,m^2)}{d x}|_{x=\zeta} +c_2 \,
\ee
where $c_1,c_2$ are arbitrary constants and $\zeta$ is some suitably chosen scale.

 Actually, owing unrenormalized result (\ref{result3}) is enough to make an explicit comparison with  calculation performed in other RSs.
Making the substitution (\ref{subst2}) and  subtracting  divergent term as suggested we can get the familiar result \cite{DSE4},
  i.e. the expression obtained via dimensional regularization followed by subtractions (or  equivalently by R-operation) establishing 
  thus BPHZ momentum scheme for which $c_1=c_2=0$. Notably, both results are  equivalent to the  Cutkosky rules method applied to the 
  sunset diagram earlier  \cite{BABEBOBU1995},\cite{BADERO2001}.

  \section{Conclusion and future prospects}

 Two novel perturbative renormalization  schemes have been introduced.
The first one,SRS, can provide the renormalized result identical to dimensional regularization and renormalization method.
The second scheme, although formally derived as a limiting case of the first one, renders the theory finite without the (infinite) counterterm required.
However, in the latter case, the momentum dependence of the radiative corrections derived is altered. 
To compare qualitatively, the effective charge has been calculated in the finite QCD. It represent the first  comparison with experimental results, providing the known value $\alpha_s(M_Z)=0.119$,  assuming approximate scheme independence at this scale.

The finite scheme, called FRRS in the main text, offers the possibility of dealing with quantum field theory correlators as perfectly finite objects.
While, in SRS some ambiguities persist, since  the fine tuned  division of regulators is required to preserve symmetry in the theory, the finite
renormalization scheme is unique. Of course, experimental limitations like precise measurment of $\alpha_{QED}$ in the spacelike region or measurment of $\alpha_{QCD}$ enough high above electroweak scale can  rule out the FRRS at all. 

While SRS can be wieved as not useful, since providing already known  DIMR result, the finite scheme can shed new light in many directions.
Obviously, the unification of effective couplings at some unification scale become an unsupported phenomena and at least the QCD effective charge remains solely frozen at relatively large value  $\alpha_{QCD}(\infty)\simeq 0.09$. Since the approach is just born, there are many ways to be improved. One can consider higher orders and then look to lower scales as well. Although the proposed method is purely perturbative, the small value 
of QCD effective charge makes a comparison with lattice method challenging. Needless to say, there is still not to much done when comparing
gauge dependent , i.e. unobservable quantities between continous (here) and lattice theory  \cite{Mass2017}.

The scheme can be promising in further respect, it  can be applied to phenomena where one needs to get rid of  the divergences from the principle.
 Chiral symmetry breaking and calculations of hadron properties in the functional formalism have already been mentioned. However, it can also be useful for dealing with naively non-renormalizable models \cite{little1,little2,little3}, where chiral perturbation theory could be possible hot candidate for application of finite scheme.

\appendix

\section{Photonic VHP in the gauge sector of the SM}

 The general form of photon self-energy of in $R_{\xi}$ gauges consist from seven distinct  one loop Feynman diagrams, the one with fermion loop 
  was discussed separately in the previous Section due to its own peculiarity.
 Each diagram can be decomposed to two terms containing  the transverse and longitudinal projector with  its own scalar form factor  function. These functions
  can be written in terms of linear combinations of  two functions $a(M_i)$ and $b(q^2,M_i,M_j)$, the former is quadratically divergent.
 All diagrams  are individually  finite for a finite gauge parameter $\xi$.  
  
  Actually, it is quite obvious that  we do not need to proceed Feynman paramaterization again and again and  
  the result can be derived by a simple algebraic manipulation, e.g. by using the identity
 $2(k.q)=[(l+q)^2-m^2)]-[l^2-m^2]-q^2$ and by shifts of integral variables.
 Thus for instance one can immediately write for the following integral:
\bea
&&i\hat{L} \int\frac{d^4 l}{(2\pi)^4} \frac{(l.q)^2}{[l^2-m^2+\ep][(l+q)^2-m^2+\ep]}=
i\hat{L} \int\frac{d^4 l}{(2\pi)^4}\left[\frac{l.q}{2[l^2-m^2+\ep]}-\frac{l.q-q^2}{2[(l+q)^2-m^2+\ep]}\right]
\nn \\  
&+&i\hat{L}\int \frac{d^4 l}{(2\pi)^4}\frac{q^2}{2[l^2-m^2+\ep][(l+q)^2-m^2+\ep]}=\frac{q^2}{2} a(m^2)+\frac{q^2}{4}b(q^2,m^2) \, .
\eea   
which will be used in the following derivation.

 In t'Hooft-Feynman gauge, which we use for simplicity here, the most dominant expression is given by diagram with two trilinear  $WW\gamma$ vertices.
 The vertices  have usual Yang-Mills structure and after some algebra the familiar contribution reads
 \be \label{up}
\Pi_{\mu\nu}(1,q)=\frac{e^2}{2} i \hat{L} \int \frac{d^4 l}{(2\pi)^4} \frac{-\lambda^2}{-\lambda^2}\frac{N_{\mu\nu}(q,l)}{[l^2-M_W^2+\ep][(l+q)^2-M_W^2+\ep]}\, ,
\ee
where 
\be
N^{\mu\nu}(q,l)= 10 l^{\mu\nu}+5(l^{\mu}q^{\nu}+l^{\nu}q^{\mu})-2 q^{\mu}q^{\nu}+g^{\mu\nu}( 2l^2+2k.q+5q^2)\, ,
\ee
 where  again the symbol $ \hat{L}$ is a formal remainder that we should perform $L$-operation with f=1, instead of direct momentum integration.
   
 Projecting (\ref{up}) with  $P_L(q)= q^{\mu}q^{\nu}/q^2$ and $P_T(q)= g^{\mu\nu}-q^{\mu}q^{\nu}/q^2$ one gets
 the contribution to the longitudinal and to the transverse part of selfenergy.
   The former  reads
 \bea
 \Pi_L(1,q^2)&=&\frac{e^2}{2} i \hat{L} \int \frac{d^4 l}{(2\pi)^4}  \frac{2l^2+12 l.q+3q^2+10\frac{(l.q)^2}{q^2}}{[l^2-M_W^2+\ep][(l+q)^2-M_W^2+\ep]}\, ,
 \nn \\
 &=&\frac{7e^2}{2} a(M_W^2,\lambda_1)+e^2[M_W^2-\frac{q^2}{4} b(q^2;M_W^2,\lambda_1)] \, .
 \eea
The tadpole diagram with WWAA quartic vertex gives the term proportional solely to metric tensor. In T-L decomposition of polarization function  it thus gives
\be
\Pi_L(2,q^2)=-3e^2 a(M_W^2,\lambda_2) \, .
\ee

Two  diagrams with two lines of charged scalar Goldsteone are purely transverse , giving thus trivial contribution to the function $\Pi_{L}$. A single one loop 
 diagram, which  involves one scalar and one vector propagator provides the following nontrivial contribution
 \be
\Pi_L(3,q^2)=-e^2 m^2_W b(q^2,M^2_W,\lambda_3) \, .
\ee

The forth and the last contribution follows from the  diagram with two ghost propagators, which reads
\be
\Pi_L(4,q^2)=-\frac{e^2}{2} a(M_W^2,\lambda_4)+e^2\frac{q^2}{4}b(q^2,M^2_W,\lambda_4) \, .
\ee

Summing all one loop diagrams for identical $\lambda$'s one finally gets:
\be
\Pi_L(q^2)=\sum_{i=1}^4\Pi_L(i,q^2)=0\, ,
\ee
which means the WTI $q^{\mu}\Pi_{\mu\nu}=0$ is satisfied in the gauge sector automatically for single choice of $\lambda$.

%%%%%%%%%%%%%%%%%%%%%%%%%%%%%%%%%%%%%%%%%%%%%%%%%%%%%%%%%%%%%%%%%%%%%%%

%

\begin{thebibliography}{00}
\bibitem{doubletad}
Appropriate diagram is figured for instance  in Eq. 5.15. of ref. \cite{RAMON}, page of  \cite{pokorski} or  page 236 fig.5.16 of textbook \cite{brown}.
\end{thebibliography}
\end{document}